\documentclass[final,3p,times]{elsarticle}

\usepackage{amssymb}

\usepackage{amsthm}
\begin{document}

\begin{frontmatter}



\title{ Random versus holographic fluctuations of the background metric. I. \\ (Cosmological) running of space-time dimension }


\author{ Michael~Maziashvili }
\ead{mishamazia@hotmail.com}

\address{Andronikashvili
Institute of Physics, 6 Tamarashvili St., Tbilisi 0177, Georgia \\

Faculty of Physics and Mathematics, Chavchavadze
State University, 32 Chavchavadze Ave., Tbilisi 0179, Georgia}

\begin{abstract}

A profound quantum-gravitational effect of space-time dimension running with respect to the size of space-time region has been discovered a few years ago through the numerical simulations of lattice quantum gravity in the framework of causal dynamical triangulation [hep-th/0505113] as well as in renormalization group approach to quantum gravity [hep-th/0508202]. Unfortunately, along these approaches the interpretation and the physical meaning of the effective change of dimension at shorter scales is not clear. The aim of this paper is twofold. First, we find that box-counting dimension in face of finite resolution of space-time (generally implied by quantum
gravity) shows a simple way how both the qualitative and the quantitative features of this effect can be understood. Second, considering two most interesting
cases of random and holographic fluctuations of the background
space, we find that it is random fluctuations that gives running dimension resulting in modification of Newton's inverse square law in a perfect agreement with the modification coming from one-loop gravitational radiative corrections.

\end{abstract}


\end{frontmatter}

\section{Introduction }

Recently a profound quantum gravitational effect of dimension
reduction of space\,-\,time near the Planck scale was discovered
in the framework of two different approaches to quantum gravity
\cite{LRAJL}. As thus far there is no final picture of quantum
gravity and the assault on it goes along several ways, it is very
desirable to derive this result from an underlaying principle that
is common for all approaches. Most likely such a fundamental
principle seems to be finite resolution of space\,-\,time, like
quantum mechanics implies finite resolution of phase space. In
this or another way, quantum gravity strongly indicates the finite
resolution of space\,-\,time, that is, space\,-\,time uncertainty.
Space\,-\,time uncertainty is common for all approaches to quantum
gravity be it: space\,-\,time uncertainty relations in string
theory \cite{String1, String2}; noncommutative space\,-\,time
approach \cite{noncommutative}; loop quantum gravity \cite{Loop};
or space-time uncertainty relations coming from a simple {\tt
Gedankenexperiments} of space\,-\,time measurement
\cite{Gedankenexperiments}. Well known entropy bounds emerging via
the merging of quantum theory and general relativity also imply
finite space\,-\,time resolution \cite{entropybounds}. The
combination of quantum theory and general relativity in one or
another way manifests that the conventional notion of distance
breaks down the latest at the Planck scale $l_P \simeq
10^{-33}$\,cm \cite{minimumlength}. Indeed, this statement can be
understood in a very simple physical terms. (In what follows we
will assume system of units $\hbar = c = k_B = 1$). Namely, posing
a question to what maximal precision can one mark a point in space
by placing there a test particle, one notices that in the
framework of quantum field theory the quantum takes up at least a
volume, $\delta x^3$, defined by its Compton wavelength $\delta x
\gtrsim 1/m$. To not collapse into a black hole, general
relativity insists the quantum on taking up a finite amount of
room defined by its gravitational radius $\delta x \gtrsim
l_P^2m$. Combining together both quantum mechanical and general
relativistic requirements one finds
\begin{equation}\label{abslimit} \delta x \gtrsim
\mbox{max}(m^{-1},~l_P^2m)~.\end{equation} From this equation one
sees that a quantum occupies at least the volume $ \sim l_P^3 $.
Since our understanding of time is tightly related to the periodic
motion along some length scale, this result implies in general an
impossibility of space\,-\,time distance measurement to a better
accuracy than $\sim l_P$. Therefore, the point in space\,-\,time
can not be marked (measured) to a better accuracy than $ \sim
l_P^4$. It is tantamount to say that the space\,-\,time point
undergoes fluctuations of the order of $\sim l_P^4$, we refer the
reader to a very readable paper of Alden Mead \cite{minimumlength}
for his discussion regarding the status of a fundamental (minimum)
length $l_P$, as this conceptual standpoint was unanimous in
almost all subsequent papers albeit many authors apparently did
not know this paper. Over the space\,-\,time region $l^4$ these
local fluctuations add up in this or another way that results in
four volume fluctuation of $l^4$. In view of the fact how the
local fluctuations of space\,-\,time add up over the macroscopic
scale ($l \gg l_P$), different scenarios come into play. Most
interesting in quantum gravity are random and holographic
fluctuations. From the very outset let us notice that the length
scale $l$ we are interested in is a horizon distance $l_H$. If the
local fluctuations, $\l_P$, are of random nature then over the
length scale $l_H$ they add up as $\delta l_H =
(l_H/l_p)^{1/2}l_P$. In the holographic case, the local
fluctuations, $\l_P$, add up over the length scale $l_H$ in such a
way to ensure the black hole entropy bound on the horizon region
$\delta l_H = (l_H/l_P)^{1/3}l_P$. Throughout this paper we will
consider these two cases separately. Taking note of finiteness of
the space-time resolution in quantum gravity, one immediately
faces the question what operational meaning can be given to the
space\,-\,time dimension. The fundamental to the generalized
mathematical treatment of dimension for a set under consideration
is an idea of measurement at scale $\epsilon$, for each $\epsilon$
we measure a set in a way that ignores irregularities of size less
than $\epsilon$, and we see how this measurement behaves as
$\epsilon \rightarrow 0$. For more details we refer the reader to
a very readable book of Falconer \cite{Falconer}. As Falconer
notices in the introduction of his book , "A glance at a recent
physics literature shows the variety of natural objects that are
described as fractals $-$ cloud boundaries, topographical
surfaces, coastlines, turbulence in fluids, and so on. None of
these are actual fractals $-$ their fractal features disappear if
they are viewed at sufficiently small scales." However, this naive
expectation is impeded by quantum gravity.

\section{ Box\,-\,counting dimension }

Because of quantum gravity the dimension of space\,-\,time appears
to depend on the size of region, it is somewhat smaller than $4$
and monotonically increases with increasing of size of the region
\cite{LRAJL}. We can account for this effect in a simple and
physically clear way that allows us to write simple analytic
expressions for space\,-\,time dimension running. In what follows
we will use a box\,-\,counting dimension \cite{Falconer}.
Box\,-\,counting dimension is one of the most widely used
dimension largely due to its ease of mathematical calculation and
empirical estimation. A major disadvantage of the Hausdorff
dimension \cite{Hausdorff} is that in many cases it is hard to
calculate or to estimate by computational methods. Except of some
"pathological" cases that have no physical interest, the Hausdorff
dimension is equivalent to the box\,-\,counting dimension
\cite{Kolmogorov}. Let us consider a set $\mathcal{F}$ that is
understood to be a subset of four dimensional Euclidean space
$\mathbb{R}^4$, and let $l^4$ be a smallest box containing this
set, $\mathcal{F} \subseteq l^4$. The mathematical concept of
dimension tells us that for estimating the dimension of
$\mathcal{F}$ we have to cover it by $\epsilon^4$ cells and
counting the minimal number of such cells, $N(\epsilon)$, we can
determine the dimension, $d \equiv \dim(\mathcal{F})$ as a limit
$d = d(\epsilon \rightarrow 0)$, where $n^{d(\epsilon)} = N $ and
$n = l/\epsilon$. For more details see \cite{Falconer}. This
definition can be written in a more familiar form as
\[ d =\lim\limits_{\epsilon \rightarrow 0} {\ln N(\epsilon) \over \ln {l \over \epsilon}}~.\]
Certainly, in the case when $\mathcal{F} = l^4$, by taking the
limit $d(\epsilon \rightarrow 0)$ we get the dimension to be $4$.
From the fact that we are talking about the dimension of a set
embedded into the four dimensional space, $\mathcal{F} \subset
\mathbb{R}^4$, it automatically follows that its dimension can not
be greater than $4$, $d \le 4$. Hence, for a fractal $\mathcal{F}$
uniformly filling the box $l^4$ we have the reduction of its
volume
\[ V(\mathcal{F}) = \lim\limits_{\epsilon \rightarrow 0} N(\epsilon)\epsilon^4
= \lim\limits_{\epsilon \rightarrow
0}n(\epsilon)^{d(\epsilon)}\epsilon^4 ~,\] in comparison with the
four dimensional value that would be \[ \lim\limits_{\epsilon
\rightarrow 0}n(\epsilon)^4\epsilon^4  = l^4~.\] Introducing
$\delta N = n(\epsilon)^4 - N(\epsilon)$, the reduction of
dimension $\varepsilon = 4 - d$ can be written as
\begin{equation}\label{opdim} \varepsilon(\epsilon) \,=\, - {\ln
\left(1 - {\delta N(\epsilon) \over n(\epsilon)^4 } \right)\over
\ln n(\epsilon) } \, \approx \, {1 \over \ln n(\epsilon)}
\,{\delta N(\epsilon) \over n(\epsilon)^4}~. \end{equation}
Quantum gravity, whatever the particular approach is, shows up a
finite space\,-\,time resolution. The local fluctuations,
$\epsilon = l_P$, add up over the length scale $l$ resulting in
fluctuation $\delta l(l)$. Respectively, for the region $l^4$ we
have the deviation (fluctuation) from the four dimensional value
of volume of the order $\delta V = \delta l(l)^4$. One naturally
finds that this fluctuation of volume has to account for the
reduction of dimension\footnote{This suggestion has been made in
\cite{mazia1}, though the rate of volume fluctuation was
overestimated in this paper. Let us also notice that the necessity
of operational definition of dimension because of quantum
mechanical uncertainties (not quantum\,-\,gravitational !) was
first stressed in \cite{ZS}.}. It is worth noticing that albeit
locally (that is, at each point) the space\,-\,time undergoes
fluctuations of the order $\sim l_P$, for the fluctuations add up
over the length scale $l$ to $\delta l(l)$, the region $l^4$
effectively looks as being made of cells $\delta l(l)^4$ that
immediately prompts the rate of volume fluctuation.

\section{ Dimension running/reduction of space\,-\,time in the case of random fluctuations }

In the case when local fluctuations of space\,-\,time are of
random nature we expect the Poison fluctuation of volume $l^4$ of
the order $ \delta V = \sqrt{l^4/l_P^4}\, l_P^4 $ \cite{Sorkin}.
Simply, this value of $\delta V$ can be understood in the
operational sense that in measuring of volume $l^4$ with the
precision $l_P^4$ one naturally expects the error $l_P^4$ to take
on $\pm$ sign with equal probability at each step of measurement
that leads to the summation of error with the factor
$\sqrt{l^4/l_P^4}$. The same can be said in the way that the local
fluctuations $l_P$ take on $\pm$ sign along the length scale $l$
with equal likelihood that results in amplification factor
$\sqrt{l/l_P}$ over this length scale, see for a detailed
discussion \cite{mazia1}. Respectively, from Eq.(\ref{opdim}) one
gets $n = l/l_P$, $\delta N = \sqrt{l^4/l_P^4} = l^2/l_P^2$,

\begin{equation}\label{opdimrandom} \varepsilon_{random} = {1 \over \ln {l\over l_P}}\, \left({l_P \over
l}\right)^2~.\end{equation} This equation gives the running of
dimension with respect to the size of region $l$.

\section{ Dimension running/reduction of space\,-\,time in the case of holographic fluctuations }

Considering a weakly gravitating system in asymptotically flat
space-time, the Bekenstein entropy bound tells us that the maximum
number of bits that can be stored inside the region $l^3$ with the
energy $E$ can not exceed \cite{entropybounds}

\begin{equation}\label{entensize} S \lesssim E\,l~.\end{equation}
We will typically ignore the numerical factors of order unity and
will make an effort to keep the equations as simple as possible in
order to not obscure the underlying physical concepts. Maximum
number of bits is set respectively by $E_{max}\sim l/l_P^2$ above
which the gravitational collapse of this energy into a black hole
will take place
\begin{equation}\label{maxent} S_{BH} \,\simeq\, \left( l \over
l_P \right)^2~.\end{equation} Taking note of this fact, that the
maximum amount of information available to an observer within the
cosmological horizon is given by Eq.(\ref{maxent}) with $l = l_H$,
one finds the maximal space-time resolution over the horizon scale
to be\footnote{Combining quantum mechanics with general
relativity, the relation $\delta l \gtrsim l_P^{2/3}l^{1/3}$ as an
intrinsic imprecision in measuring of length scale $l$ for the
Minkowskian background  was obtained by K\'arolyh\'azy in 1966
\cite{Gedankenexperiments}.}
\begin{equation}\label{Karolunce}
  \delta l_H \, =  \, {l_H \over
  S_{BH}^{1/3}} \,\simeq\, l_P^{2/3}l_H^{1/3}~. \end{equation}
  Thus in the holographic case the four volume $V = l_H^4$ undergoes fluctuation of the order
  $\delta V =   \delta l_H ^4 \simeq l_P^{8/3}l_H^{4/3} $ that
  with respect to the Eq.(\ref{opdim}) yields

\begin{equation}\label{opdimholographic} \varepsilon_{holographic}= {1 \over \ln {l_H\over l_P}}\, {\delta V \over V}
= {1 \over \ln {l_H\over l_P}}\, \left({l_P \over
l_H}\right)^{8/3}~.\end{equation} It is curious to notice that if
one assumes the holographic fluctuations to pertain to the space
only but not to the time, that is, if we use three volume instead
of the four one in Eq.(\ref{opdimholographic}), the dimension will
coincide with (\ref{opdimrandom}). This convergence of results
seems intriguing, so one could simply argue the use of three
volume instead of the four one because the entropy bound
Eq.(\ref{entensize}) has to do immediately with the spatial
region, but it is certainly a bit subtle question needing further
scrutiny.

\section{ QFT reasoning for understanding of space\,-\,time dimension reduction in light of quantum gravity  }

It is an old well known idea that the melding of quantum theory
and gravity typically indicates the presence of an inherent UV
cutoff. In view of the above discussion, the emergence of such an
intrinsic UV scale can be understood in a simple physical way that
the background metric fluctuations does not allow QFT to operate
with a better precision than the background space resolution. That
is, if we have a characteristic IR scale $l$, then the UV cutoff,
$\Lambda$, is naturally bounded by the fluctuation $\delta l(l)$,
$\Lambda \lesssim  1/\delta l(l)$. In its turn, the presence of IR
scale is well motivated by the existence of a cosmological
horizon, $l \lesssim l_H$. Thus knowing a particular IR scale, the
presence of corresponding UV cutoff tells us that the Feynman
diagrams pertaining to this theory become finite. This result in
terms of dimensional regularization inevitably favors the
dwindling of dimension.

\section{ Discussion }

First of all let us discuss the validity region for
Eqs.(\ref{opdimrandom},\,\ref{opdimholographic}). From the above
discussion one simply infers that the validity condition is $\varepsilon \ll 1$. That is, the discussion is valid as long as
four volume fluctuation $\delta V$ satisfies $\delta V \ll V =
l^4$. How far in the early cosmology can we use
Eqs.(\ref{opdimrandom},\,\ref{opdimholographic})\,? Say, for the
length scale $\sim 1/ 10^{16}\mbox{GeV}$ corresponding to the GUT,
the $\varepsilon_{random}$ that is larger than
$\varepsilon_{holographic}$ gives $\varepsilon_{random} \sim
10^{-6}$, so that the validity condition is satisfied with good
accuracy. Recalling that the inflation energy scale,
$E_{inflation}$, is bounded from above by (non) observation of
tensor fluctuations of the cosmic microwave background radiation
(relict gravitational wave background) \cite{RubSazhVery}, with
the current limit being $E_{inflation} \lesssim 10^{16}$\,GeV
\cite{Spergel}, one infers that even during the inflation stage we
can safely use the
Eqs.(\ref{opdimrandom},\,\ref{opdimholographic}).

It is important to decide between the Eqs.(\ref{opdimrandom},\,\ref{opdimholographic}) for both of them can not be true. In the low energy regime ($\ll m_P$) general relativity can be
successfully treated as an effective quantum field theory
\cite{Donoghue}. So that it is possible to unambiguously compute
quantum effects due to graviton loops, as long as the momentum of
the particles in the loops is cut off at some scale $\ll m_P$. The
results are independent of the structure of any ultraviolet
completion, and therefore constitute genuine low energy
predictions of any quantum theory of gravity. Following this way
of reasoning it has been possible to compute one-loop quantum
correction to the Newtonian potential \cite{Khriplovich}

\begin{equation}\label{oneloop}V(r) = -l_P^2{m_1\,m_2 \over r}\left[1 + \frac{41}{10\pi}  \left({l_P \over r}\right)^2 \right]
~.\end{equation} Let us compare this result with the modification of the Newton's law due to dimension running Eqs.(\ref{opdimrandom},\,\ref{opdimholographic}).

Modification of the Newton's law due to dimension reduction can be
estimated without too much trouble by writing it in the Planck
units

\[ -l_Pm_1\,m_2 \,{ l_P \over r} \equiv  -l_Pm_1\,m_2 \,{ 1 \over \xi} ~. \] One easily finds

\[ { 1 \over \xi^{1 - \varepsilon}}  = {e^{\varepsilon\ln\xi} \over \xi} =  {1 \over \xi} \left(1 +  \varepsilon\ln\xi +
\ldots~\right)~.\] Substituting the Eq.(\ref{opdimrandom})
\[ \varepsilon = {1 \over \xi^2\ln\xi} ~,\] we get

\[V(r) = -l_P^2{m_1\,m_2 \over r}\left[1 +  \left({l_P \over r}\right)^2 \right] ~,
\] which appears to be in perfect agreement with the Eq.(\ref{oneloop}). While the Eq.(\ref{opdimholographic}) gives obviously incorrect result.

Thus the
model of random fluctuations appears to be favored over the
holographic one. It is interesting to notice that dark energy models based on the space-time uncertainty relations also manifest the random fluctuations to be more likely \cite{darken}. Curiously enough, the uncertainty relation
$\delta l = (l_Pl)^{1/2}$ is favored over the K\'arolyh\'azy
uncertainty relation, $\delta l = l_P^{2/3}l^{1/3}$, even in the
framework of {\tt Gedankenexperiments} for space-time measurement
\cite{mazia1}.

It is somewhat
disappointing that hitherto we do not know how to work at a
fundamental level with the theories having dynamical dimension.
Nevertheless, some attempts to study the cosmology with a variable
space dimension have been already made in literature, see for
instance \cite{Mansouri}. No doubt it would be very interesting to
take a close look at the cosmology with the running dimension in
order to identify the corresponding experimental signatures.
Besides the early cosmology, for a phenomenological study of
quantum\,-\,gravitational reduction/running of space\,-\,time
dimension, the QFT effects measured with a high precision call for
attention for one can estimate in a systematic way the
corresponding quantum corrections. Such an investigation for
studying the influence of the dimension running on the running of
gauge couplings has been done in our paper \cite{mazia2}. In an
upcoming paper \cite{MS} we studied the corrections to the
hydrogen spectrum due to dimension reduction. A few experimental
signatures of this kind can be found in \cite{ZS, expsignature}.

\section*{ Acknowledgments }

 Useful comments from Zurab~Silagadze are acknowledged. This work was supported in part by the \emph{CRDF/GRDF} grant and the
\emph{Georgian President Fellowship for Young Scientists}.

\end{document}